\documentclass[aps,pre,reprint,groupedaddress,longbibliography]{revtex4-1}
\usepackage{amsmath}
\usepackage{parskip}
\usepackage{enumitem}
\usepackage{amssymb}
\usepackage{overpic}
\usepackage{hyperref}
\usepackage{soul}
\usepackage[makeroom]{cancel}

\usepackage{float,graphicx}
\usepackage[usenames, dvipsnames]{color}

\hypersetup{
	colorlinks,
	citecolor=blue,      
	filecolor=red,
	linkcolor=blue,
	urlcolor=blue,
	hyperfigures
}

\def\XXint#1#2#3{{\setbox0=\hbox{$#1{#2#3}{\int}$}
     \vcenter{\hbox{$#2#3$}}\kern-.5\wd0}}

\usepackage[page]{appendix}

\graphicspath{{./Figures/}}

\begin{document}

\title{Dispersion of run-and-tumble microswimmers through disordered media}
\author{David Saintillan}\email[\vspace{-0.6cm}Email address: ]{dstn@ucsd.edu}
\affiliation{Department of Mechanical and Aerospace Engineering, University of California San Diego, 9500 Gilman Drive, La Jolla, CA 92093, USA}


\begin{abstract}
Understanding the transport properties of microorganisms and self-propelled particles in porous media has important implications for human health as well as microbial ecology.\ In free space, most microswimmers perform diffusive random walks as a result of the interplay of self-propulsion and orientation decorrelation mechanisms such as run-and-tumble dynamics or rotational diffusion. In an unstructured porous medium, collisions with the microstructure result in a decrease in the effective spatial diffusivity of the particles from its free-space value.\ Here, we analyze this problem for a simple model system consisting of non-interacting point particles performing run-and-tumble dynamics through a two-dimensional disordered medium composed of a random distribution of circular obstacles, in the absence of Brownian diffusion or hydrodynamic interactions. The particles are assumed to collide with the obstacles as hard spheres and subsequently slide on the obstacle surface with no frictional resistance while maintaining their orientation, until they either escape or tumble.\ We show that the variations in the long-time diffusivity can be described by a universal dimensionless hindrance function $f(\phi,\mathrm{Pe})$ of the obstacle area fraction $\phi$ and P\'eclet number $\mathrm{Pe}$, or ratio of the swimmer run length to the obstacle size. We analytically derive an asymptotic expression for the hindrance function valid for dilute media ($\mathrm{Pe}\,\phi\ll 1$), and its extension to denser media is obtained using stochastic simulations.
\end{abstract}

\maketitle

\section{Introduction}

Self-propelled particles, from motile microorganisms to synthetic microswimmers, perform random walks in space that allow them to explore their environment, for instance in their quest for oxygen or nutrients. These random dynamics result from the interplay of self-propulsion and orientational fluctuations, which cause stochastic changes in their swimming direction. One classic example is the case of run-and-tumble bacteria, which perform straight runs in a given direction alternating with random reorientation events known as tumbles that are driven by the rapid unbundling and rebundling of their flagella. As first explained by Berg \cite{B1993}, the resulting random walks lead to diffusive spreading at long times, with a mean squared displacement growing linearly with time as $\langle |\Delta \mathbf{r}|^2\rangle\sim 2d D_0 t$, where $d$ is the spatial dimension and $D_0$ is an effective diffusivity. Under the assumptions of instantaneous and uncorrelated tumbles and of exponentially distributed run times, a simple random walk model predicts $D_0=v_0^2 \overline{\tau}/3$, where $v_0$ and $\overline{\tau}$ are the constant run speed and mean run time, respectively.\ These stochastic dynamics play a key role in various transport strategies such as chemotaxis, where bacteria can bias their tumbling frequency based on the local concentration of a chemical, resulting in a net drift along the chemical gradient.\ While synthetic microswimmers do not perform run-and-tumble dynamics, they typically experience rotational Brownian motion, which also leads to correlated random walks and diffusive spreading on long time scales~\cite{HJRGVG2007}. 

Motile bacteria and other microorganisms often reside in complex environments such as soils or tissues, where their frequent interactions and collisions with the microstructure strongly affect their motions.\ Understanding active dispersion in such systems is key to a variety of problems in soil ecology, biofouling and bioremediation, as well as in medicine where it affects the spread of bacterial infections. Additionally, the potential of engineered active particles lies in their ability to navigate complex geometries, be it in lab-on-a-chip devices or inside living organisms for drug-delivery applications.\ Our fundamental understanding of basic transport properties of active particles in heterogeneous random media remains, however, incomplete \cite{BDLRVV2016}. 

Recent microfluidic experiments using either living microorganisms or synthetic self-propelled particles have started to shed light on the physics of active transport in these complex environments \cite{MTD2022,KSA2022}.\ The ability to fabricate model porous media of controlled porosity and microstructure provides a useful tool for probing the role of geometry and crowding in determining long-time dispersion. In both random \cite{BVDVSLP2016,SNDG2015,LMFS2016,SSS2017,FVMSSGD2019,BD2019,BD2019b,DMPBD2020,MXT2022} and periodic \cite{BBCPR2019,DWDG2019,RHBTB2015,WGS2021,CQGL2022,DWG2023} media, the leading effect of the porous microstructure is to hinder particle transport as a result of frequent collisions between microswimmers and obstacles, resulting in a decrease in the effective diffusivity with the volume fraction of the medium.\ While the precise nature of the scattering dynamics occurring at obstacles is found to depend on the type of microswimmer \cite{KDPG2013,CLTKP2015,MBACV2019,DGCHSVGE2015,OSBKCMB2018,MCCB2017} and potential role of hydrodynamic interactions \cite{TPBSG2014,SMBL2015}, all self-propelled particles in confinement have a tendency to accumulate at boundaries \cite{BTBL2008,BSD2017,ES2015,EAS2015,YB2015,SZS2015}, with the effect of reducing their run length thereby impeding transport. In strongly confined environments (low-porosity media), motile bacteria have even been observed to abandon run-and-tumble dynamics in favor of other more efficient transport strategies \cite{BD2019,BD2019b}.\ The role of obstacle shape has also been considered, with asymmetric obstacles potentially giving rise to rectified motion \cite{DZTALRWSZ2017}.\ Finally, a few experiments have considered the role of an externally applied flow \cite{CCDDA2019,DWDG2019}, which has a strong effect on mean transport and dispersion by reorienting the swimmers in the fluid shear generated by the microstructure \cite{AMPRLRC2013,SVMKERS2020,SMPBA2020}.  

Modeling efforts aimed at predicting dispersion in complex media have been more limited, due in part to challenges in accounting for details of the scattering dynamics and porous medium geometry.\ On the computational side, various numerical simulations have been performed based on the active Brownian particle (ABP) model in porous media described as random distributions of obstacles \cite{ZWS2017,CP2013,MXT2022} as well as in periodic post arrays \cite{PFF2014,JCB2019,ACS2019}, including in the presence of hydrodynamic interactions \cite{CIL2017}.\ Analytical predictions, however, have been very scarce with a few exceptions.\ Theoretical models have been proposed for transport of active particles in cubic lattices in the presence of obstacles \cite{BZBTV2018,RSBI2022}: while these models allow for analytical predictions, their underlying assumptions make them difficult to compare with real systems.\ In periodic geometries, generalized Taylor dispersion theory has been applied to estimate effective transport coefficients such as the mean velocity and long-time swim diffusivity of ABPs \cite{ACS2019}.\ Very recently, the case of random media was also addressed using a continuous random walk approach modeling the effect of interactions with the porous microstructure as random trapping events \cite{DCDCA2022}.\ Yet, a general theoretical framework able to yield closed-form expressions for the diffusivity in a random medium remains lacking, even under the most basic assumptions. 

Here, we propose a minimal theoretical model for the dispersion of microswimmers through a disordered medium.\ We consider point-like run-and-tumble micro-swimmers traveling in two dimensions through the interstices of a random distribution of circular obstacles in the absence of Brownian diffusion or hydrodynamic interactions. Simple interaction rules are adopted whereby a swimmer colliding with an obstacle simply slides on its surface without friction while maintaining its orientation, until it either tumbles or escapes by swimming away tangentially to the surface.\ A related model was proposed by Jakuszeit \textit{et al.}~\cite{JCB2019} to analyze transport through periodic arrays; we apply it to the case of random disordered media.\ As we show below, the effect of collisions with the microstructure on the diffusivity can be captured by a dimensionless hindrance function $f(\mathrm{Pe},\phi)$, which is a function of the P\'eclet number $\mathrm{Pe}=v_0\overline{\tau}/{a}$, or ratio of the mean run length $v_0\overline{\tau}$ to the obstacle radius ${a}$, and of the mean area fraction $\phi$ of the obstacles.\ The objective of the paper is to determine $f$, which we calculate analytically in the dilute limit defined as $\mathrm{Pe}\,\phi\ll 1$, and numerically for arbitrary values of $\mathrm{Pe}$ and $\phi$.\ The paper is organized as follows.\ Details of the problem formulation and diffusivity calculation are provided in Sec.~\ref{sec:probdef} and \ref{sec:diff}, respectively.\ The limit of dilute media is analyzed theoretically in Sec.~\ref{sec:dilutetheory}, and results from the theory are discussed and compared to numerical simulations with varying porosities in Sec.~\ref{sec:simulations}. We conclude in Sec.~\ref{sec:conclusion}.  

\section{Problem definition\label{sec:probdef}}

We analyze the dispersion of non-interacting run-and-tumble microswimmers traveling through the interstices of a random porous medium in two dimensions.\ The medium is composed of identical non-overlapping circular pillars of radius $a$, with area fraction $\phi=N_t\pi  a^2/L^2$ where $L$ is the linear dimension of the square domain and $N_t$ is the total number of pillars. The assumption of identical pillars is convenient for theoretical analysis but will be relaxed in some of the simulations of Sec.~\ref{sec:simulations}. The system is assumed to be large enough that swimmers remain far away from any domain boundaries at all times; in simulations, we will make use of periodic boundary conditions. 

\begin{figure}[b]\vspace{-0.3cm}
\includegraphics[width=0.85\columnwidth]{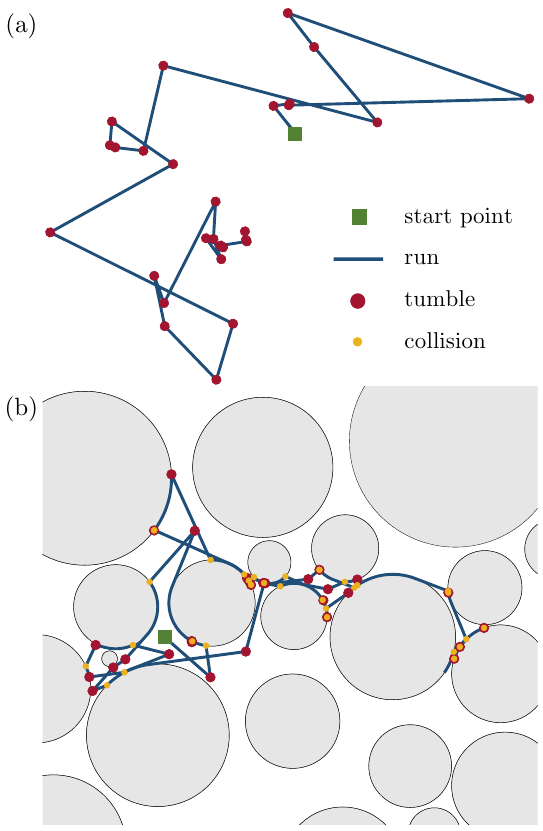}\vspace{-0.2cm}
\caption{Typical trajectories of run-and-tumble particles in free space (a) and in a two-dimensional porous medium (b), for a duration of 30 runs.\ In each case, the run time is exponentially distributed and pre- and post-tumble orientations are uncorrelated. In (b), the porous medium has a pillar area fraction of $\phi=0.62$ with a random Gaussian distribution of pillar radii with standard deviation $\sigma_a/\overline{a}=0.5$, and the P\'eclet number based on the mean radius is $\mathrm{Pe}=\overline{\ell}/\overline{a}= 2.0$. } \label{fig:RTPtrajs} \vspace{-0.0cm}
\end{figure}

In free space (no pillars), the microswimmers perform simple run-and-tumble dynamics as depicted in Fig.~\ref{fig:RTPtrajs}(a): straight runs with constant velocity $v_0$ and run time $\tau$ alternate with instantaneous reorientation events.\ The run time is a random variable governed by a probability density function $p(\tau)$ with mean value $\overline{\tau}$.\ We will consider two cases:
\begin{equation}
p(\tau)=\begin{cases}
\delta(\tau-\overline{\tau}) & \text{constant run time,} \vspace{0.1cm}\\
\overline{\tau}^{-1}\exp(-\tau/\overline{\tau}) & \text{exponential distribution.}\hspace{-0.2cm}
\end{cases} \label{eq:taudist}
\end{equation} 
The exponential distribution provides a good approximation to the distribution of run times for \textit{E. coli} \cite{BSB1983} and has been widely used in models of bacterial run-and-tumble. More detailed measurements, however, have shown deviations from the exponential model \cite{KEPC2006} and have highlighted strong temporal variability in single cells \cite{FSJDDMLC2019,FRSLAC2020}; we neglect these effects here.\ Given $v_0$ and $\tau$, we  define the run length  $\ell=v_0 \tau$, or distance traveled by the swimmer between two tumbles in the absence of pillars, with mean value $\overline{\ell}=v_0\overline{\tau}$.

\begin{figure}[t]
\includegraphics[width=0.85\columnwidth]{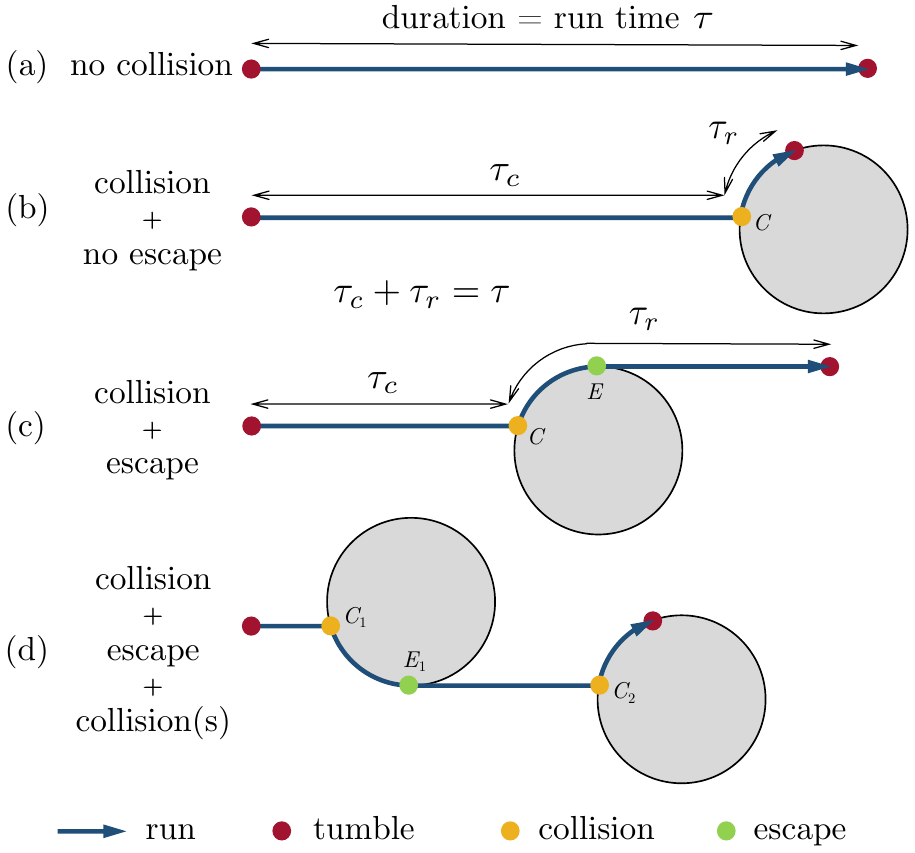}\vspace{-0.3cm}
\caption{Types of possible displacements during a single run of total duration $\tau$. If the swimmer collides with a pillar (point $C$), it can either escape (point $E$) or end its run on the pillar. The time to collision is $\tau_c$, whereas $\tau_r=\tau-\tau_c$ is the remaining time in the run after collision.  } \label{fig:runtypes} \vspace{-0.3cm}
\end{figure}

In a porous medium [Fig.~\ref{fig:RTPtrajs}(b)], microswimmers can collide with pillars, and these collisions alter their trajectories leading to scattering. We propose a minimal model for collisions based on the following assumptions:
\begin{enumerate}[label=(\roman*)]
\itemsep0em 
\item Swimmers are point particles that interact with pillars via a hard-sphere potential.
\item When a swimmer collides with a pillar, its orientation and run time remain unchanged.
\item After impact, the swimmer slides along the pillar surface with the tangential component of its swimming velocity, and no resistance to sliding. 
\item If the swimmer's orientation becomes tangent to the surface, it escapes from the pillar and continues its run in a straight line, possibly encountering additional pillars before the end of the run.
\item If the run time elapses before the swimmer is able to escape, the run ends on the pillar surface where the next tumble takes place.
\end{enumerate}
The four types of runs (no collision, collision with no escape, collision with escape, and multiple collisions) are depicted graphically in Fig.~\ref{fig:runtypes}. When a collision occurs, we denote by $\tau_c$ the time to collision from the start of the run, and by $\tau_r$ the remaining time in the run after collision, so that $\tau_c+\tau_r=\tau$. Runs with multiple collisions can be recursively modeled as sequences of single-collision runs with reduced run times.\ Any of the runs depicted in Fig.~\ref{fig:runtypes} can either start with the swimmer in the bulk or on the surface of a pillar.\ Note that $\tau_c=0$ in cases where a run starts on the surface of a pillar with the swimmer pointing into the pillar.

While $\tau$ is assumed to be unaffected by collisions, note that the actual distance traveled by a swimmer colliding with a pillar is in fact shorter than $v_0 \tau$.\ In this case, we will continue to use the variable $\ell$ to denote the \textit{unimpeded} run length $v_0\tau$.\ In a porous medium, system properties are entirely governed by two dimensionless numbers: the area fraction $\phi$ introduced above, as well as $\mathrm{Pe}=v_0\overline{\tau}/a=\overline{\ell}/a$, which compares the persistence length of swimming trajectories to the pillar size and can be interpreted as a swimming P\'eclet number.

The assumptions made here greatly idealize the dynamics of real microswimmers near walls, which are usually more complex.\ In particular, assumptions (i)--(iv) are incompatible with hydrodynamic interactions, which can lead to a long-ranged coupling between swimmers and pillars and reorient swimmers during collisions as seen in various experiments \cite{DDCGG2011,TPBSG2014,CLTKP2015,SNDG2015,BULWRZWS2019} and models \cite{SMBL2015,LKG2017}. In experimental systems, other effects can also impact orientation dynamics, including direct steric contacts especially in the case of flagellated swimmers \cite{KDPG2013,CLTKP2015,LKG2017} and rodlike swimmers \cite{BULWRZWS2019}, as well as chemical interactions in the case of phoretic swimmers \cite{DGCHSVGE2015,BVDVSLP2016,SKUPTS2016,PUDD2018}. This reorientation at boundaries in turn leads to scattering at angles that are non-tangent with the surface. The assumption of frictionless sliding is also an approximation, as either lubrication layers or surface roughness would come into play and affect tangential motion in experiments.\ Nevertheless, this minimal model provides a simple baseline for understanding the effect of collisions on average transport properties. 

\section{Diffusivity\label{sec:diff}}

As they travel through the medium, perform tumbles and collide with pillars, the microswimmers execute random walks leading to a diffusive behavior at long times \cite{B1993}.  We denote by $\mathbf{r}_0$ the position of a swimmer at $t=0$, assumed to coincide with a tumble, and by $\mathbf{r}_N$ the location of its $N$th tumble at time $t_N$:
\begin{equation}
\mathbf{r}_N=\mathbf{r}_0+\sum_{i=1}^{N}\Delta \mathbf{r}_i, \qquad t_N=\sum_{i=1}^{N}\tau_i. \label{eq:position}
\end{equation}   
At the start of run $i$, the swimmer selects a new run time $\tau_i$ following the distribution of Eq.~(\ref{eq:taudist}), and assumes a new random orientation $\mathbf{p}_i=[\cos\theta_i,\sin\theta_i]$ where $\theta_i\in[0,2\pi)$ follows a uniform distribution.\ The displacement $\Delta \mathbf{r}_i=\mathbf{r}_i-\mathbf{r}_{i-1}$ during step $i$ is a random variable expressed as
\begin{align}
\Delta\mathbf{r}_i &= v_0\tau_i \,\mathbf{p}_i+\delta\mathbf{r}_{i}, \label{eq:Deltari} \\
&=(v_0\tau_i+\delta r_i^{\parallel})\mathbf{p}_i+\delta r_{i}^{\perp}\mathbf{p}_{i}^{\perp}. \label{eq:Deltari2}
\end{align}
Here, $v_0\tau_i\,\mathbf{p}_i$ denotes the displacement in the absence of any collision. If one or more collision(s) take place during the run, this displacement is modified by a correction $\delta\mathbf{r}_{i}$, which is decomposed into longitudinal (along $\mathbf{p}_i$) and transverse (perpendicular to $\mathbf{p}_i$) contributions in Eq.~(\ref{eq:Deltari2}), where $\mathbf{p}_{i}^{\perp}=[-\sin\theta_i,\cos\theta_i]$.\ The displacements $\smash{\delta r_{i}^{\parallel}}$ and $\smash{\delta r_{i}^{\perp}}$ are random variables that depend on the collision incidence angle $\alpha$ (to be defined more precisely later) and collision time $\tau_c$, in addition to $v_0$, $\tau_i$ and $a$.\ We explain their calculation in detail in Sec.~\ref{sec:dilutetheory}.

Given Eq.~(\ref{eq:position}), we can estimate the mean squared displacement after $N$ runs as
\begin{equation}
\langle|\mathbf{r}_N-\mathbf{r}_0|^2 \rangle = \sum_{i=1}^{N}\sum_{j=1}^{N}\langle \Delta \mathbf{r}_i\cdot \Delta \mathbf{r}_j\rangle,
\end{equation}
where brackets $\langle\cdot\rangle$ denote an ensemble average over all possible run outcomes (random variables $\tau_i$, $\mathbf{p}_i$, as well as $\alpha$ and $\tau_c$ for any collisions). Assuming successive runs are uncorrelated and using Eq.~(\ref{eq:Deltari2}), we obtain
\begin{equation}
\langle|\mathbf{r}_N-\mathbf{r}_0|^2 \rangle = N \langle(v_0 \tau)^2 + 2 v_0\tau\,\delta r_\parallel+\delta r_{\parallel}^2+\delta r_\perp^2\rangle. \label{eq:MSD}
\end{equation}
At long times, the mean squared displacement grows linearly, allowing us to define the effective diffusivity $D$ as
\begin{equation}
D=\lim_{N\rightarrow\infty }\frac{1}{4}\frac{\langle|\mathbf{r}_N-\mathbf{r}_0|^2 \rangle}{\langle t_N\rangle},
\end{equation}
i.e., using Eq.~(\ref{eq:MSD}) and $\langle t_N\rangle=N \overline{\tau}$, 
\begin{equation}
D=\frac{v_0^2\langle\tau^2\rangle}{4\overline{\tau}}+\frac{2 v_0\langle\tau \delta r_{\parallel}\rangle + \langle\delta r_{\parallel}^2+\delta r_\perp^2\rangle }{4\overline{\tau}}. \label{eq:swimdiff2}
\end{equation}
In free space (no collisions, $\delta\mathbf{r}_i=\mathbf{0}$), this expression reduces to the well known value \cite{B1993}
\begin{equation}
D_0=\frac{v_0^2 \langle\tau^2\rangle}{4\overline{\tau}}=\begin{cases}
\displaystyle \tfrac{1}{4}v_0^2\overline{\tau} & \text{constant run time,} \vspace{0.1cm}\\
\displaystyle \tfrac{1}{3}v_0^2 \overline{\tau} & \text{exponential distribution.}
\end{cases}
\end{equation}
We can then rewrite the diffusivity of Eq.~(\ref{eq:swimdiff2}) as
\begin{equation}
D=D_0[1-f(\mathrm{Pe},\phi)],
\end{equation}
where the expected decrease in diffusivity due to collisions with pillars is entirely captured by a dimensionless hindrance function
\begin{equation}
f(\mathrm{Pe},\phi)=-\frac{2 v_0\langle\tau \delta r_{\parallel}\rangle + \langle\delta r_{\parallel}^2+\delta r_\perp^2\rangle }{v_0^2 \langle\tau^2\rangle}. \label{eq:hindrance}
\end{equation}
The main of objective of this paper is to determine the function $f(\mathrm{Pe},\phi)$ governing the dependence of the diffusivity on P\'eclet number and area fraction.\ We first present a theoretical model for $f(\mathrm{Pe},\phi)$ in dilute media in Sec.~\ref{sec:dilutetheory}, and generalize it to the case of arbitrary area fractions using stochastic simulations in Sec.~\ref{sec:simulations}.

\section{Theory for dilute media\label{sec:dilutetheory}}

\subsection{Collision probabilities and time to collision}

We develop an asymptotic theory for the hindrance function $f(\mathrm{Pe},\phi)$ valid in dilute media where collisions are rare.\ In this section, we assume that the pillar size $a$ is uniform and that the run length $\tau$ is constant; these assumptions will be relaxed in the numerical simulations of Sec.~\ref{sec:simulations}. For the sake of discussion, we first analyze a single run and seek to estimate the probability that a swimmer will collide with at least one pillar during that run. As mentioned in Sec.~\ref{sec:probdef}, a run can either start with the swimmer pointing into the bulk, or with the swimmer on a pillar and pointing towards its surface. For reasons that will become clear later, we need to treat these two cases separately as they have distinct collision probabilities and distinct probability density functions for the incidence angle $\alpha$. \vspace{-0.15cm}

\subsubsection{Collision of type A: $\tau_c>0$}

\begin{figure}[t]
\includegraphics[width=0.97\columnwidth]{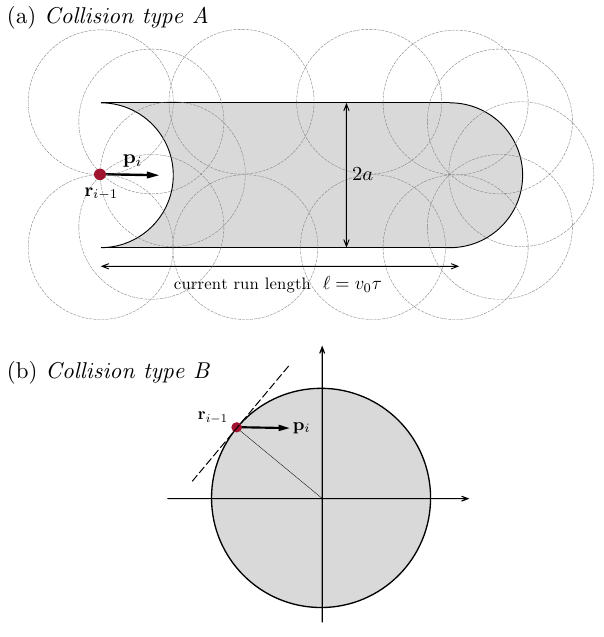} \vspace{-0.0cm}
\caption{(a) Collision of type A: for a swimmer initially pointing into the bulk, a collision will occur if the shaded region, of area $2a\ell$, contains at least one pillar. Dotted circles show the envelope of pillar positions with which a collision can occur. (b) Collision of type B: a swimmer performing a tumble on the surface of a pillar such that its new orientation points into the pillar will start its new run with a collision.} \label{fig:collision} \vspace{-0.1cm}
\end{figure}

We denote by type A a collision that occurs during a run that started with a swimmer pointing into the bulk. Note that as long as the swimmer points into the bulk, it is irrelevant whether its initial position is actually in the bulk or on the surface of a pillar. Since the initial part of the run will take place in the bulk, any collision of type A will have a strictly positive collision time $\tau_c>0$. The probability for a collision of type A to occur in any given run can be estimated graphically as shown in 
Fig.~\ref{fig:collision}(a): given that the swimmer points into the bulk, at least one pillar should have its center inside the shaded region with area $2a\ell$.\ In sufficiently dilute media, pillars are distributed randomly inside that region according to Poisson statistics.\ For a given pillar number density $n=\phi/\pi a^2$, the mean number of pillars inside the shaded region is \vspace{-0.15cm}
\begin{equation}
\langle N \rangle = 2a \ell n=\frac{2}{\pi}\frac{\ell}{a}\phi=\frac{2}{\pi}\mathrm{Pe}\,\phi. \vspace{-0.15cm}
\end{equation}
The probability $P^{c}_\mathrm{A}$ for a collision of type A is then estimated as the probability of there being at least one pillar inside the collision region: \vspace{-0.15cm}
\begin{align}
P^c_\mathrm{A}=P(N\ge 1)=1-\exp\left(-\langle N\rangle\right). \label{eq:Pa} \vspace{-0.15cm}
\end{align}
Expanding for $\mathrm{Pe}\,\phi\ll 1$,  \vspace{-0.15cm}
\begin{equation}
P^c_\mathrm{A}\approx \langle N\rangle =\frac{2}{\pi}\mathrm{Pe}\,\phi. \label{eq:Pcoldilute} \vspace{-0.15cm}
\end{equation}
In the theoretical analysis presented here, we will assume that no more than one collision can occur during a given run. To quantify the validity of this assumption, we can estimate the probability of there being two or more pillars inside the collision area: \vspace{-0.15cm}
\begin{align}
\begin{split}
P(N\ge 2) &= 1-P(N=0)-P(N=1) \\
&=1-\exp(-\langle N \rangle)-\langle N \rangle \exp(-\langle N \rangle) \\
&\approx \langle N \rangle^2. 
\end{split}
\end{align}
The assumption of no more than one collision per run is therefore valid so long as $\mathrm{Pe}\,\phi=(\ell/a)\phi \ll 1$.  Note that this condition involves the current run length $\ell$ in addition to the pillar area fraction: a swimmer might collide with multiple pillars even in dilute media if its run length is very long. Note that, in the case where $\tau$ is exponentially distributed, events will inevitably occur for which the run time is long enough that the assumption of no more than one collision breaks down.\ This effect will be quantified more precisely in the simulations of Sec.~\ref{sec:polydisp}.

Assuming a collision takes place, whether the swimmer ends its run on the pillar or is able to escape depends on the time $\tau_r$ remaining in the run after impact.\ We recall that $\tau_r=\tau-\tau_c$, where $\tau$ is the current run time and $\tau_c$ is the time to collision. For a given value of $\tau$, the location of the pillar is uniformly distributed in the shaded region of Fig.~\ref{fig:collision}(a), which implies a uniform distribution for the collision time:
\begin{equation}
p_\mathrm{A}(\tau_c)=\frac{1}{\tau},\qquad \tau_c\in(0,\tau]. 
\end{equation}
Since $\tau_r=\tau-\tau_c$, the remaining time after collision follows the same distribution: 
\begin{equation}
p_\mathrm{A}(\tau_r)=\frac{1}{\tau},\qquad \tau_r\in[0,\tau). \label{eq:taurdistA} 
\end{equation}

\subsubsection{Collision of type B: $\tau_c=0$}

A collision of type B is defined as an event where the swimmer begins its run on the surface of a pillar with a new post-tumble orientation that points into the pillar [Fig.~\ref{fig:collision}(b)]. For a collision of type B to occur, the previous run must have involved a collision (of either type A or B) in which the swimmer did not escape the pillar and thus ended its run on the surface. In that case, the new run starts with a collision with $\tau_c=0$. Estimating the probability $P^c_\mathrm{B}$ for a collision of type B is slightly more subtle, as it involves information about the previous run. We can obtain it as
\begin{equation}
P^c_\mathrm{B}=\frac{1}{2}\left[P^c_\mathrm{A}(1-P^{esc}_\mathrm{A})+P^c_\mathrm{B}(1-P^{esc}_\mathrm{B})\right], \label{eq:Pb}
\end{equation}
where $P^{esc}_\mathrm{A}$ and $P^{esc}_\mathrm{B}$ denote the probabilities of a swimmer escaping the pillar before the end of its run during a collision of either type A or B; the calculation of these probabilities involves consideration of the dynamics during collision and is deferred to Sec.~\ref{sec:escape}. The factor of $1/2$ in Eq.~(\ref{eq:Pb}) comes from the fact that a swimmer tumbling on the surface of a pillar has equal probabilities of selecting a new orientation pointing into the pillar (leading to a collision of type B) or into the bulk. Solving for $P^c_\mathrm{B}$ in Eq.~(\ref{eq:Pb}) yields
\begin{equation}
P^c_\mathrm{B}=\left(\frac{1-P^{esc}_\mathrm{A}}{1+P^{esc}_\mathrm{B}}\right)P^c_\mathrm{A}, \label{eq:Pb2}
\end{equation}
where $P^c_\mathrm{A}$ was obtained in Eq.~(\ref{eq:Pa}).

Since collisions of type B are such that $\tau_c=0$, the corresponding probability density functions for the collision and remaining times are trivial:
\begin{equation}
p_\mathrm{B}(\tau_c)=\delta(\tau_c), \qquad p_\mathrm{B}(\tau_r)=\delta(\tau_r-\tau). \label{eq:taurdistB} 
\end{equation}

\subsection{Dynamics during collision \label{sec:dyncol}}

\begin{figure}[t]
\includegraphics[width=0.7\columnwidth]{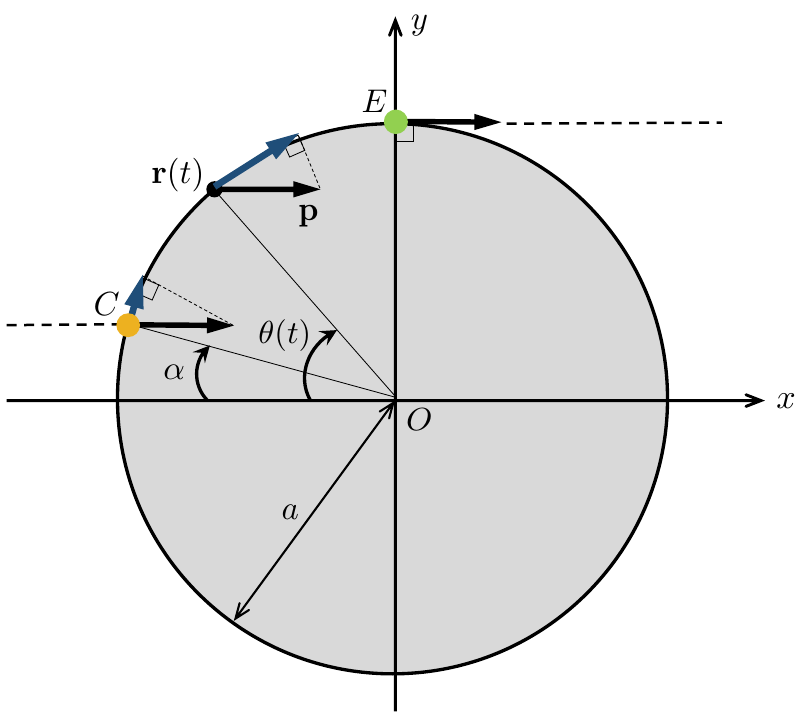} \vspace{-0.0cm}
\caption{Collision dynamics: we choose a Cartesian coordinate system as shown, with the $x$ direction aligned with $\mathbf{p}$.\ The swimmer collides at point $C$ (incidence angle $\alpha$) and slides on the surface of the pillar according to the projection of $\mathbf{p}$ in the tangent direction (blue arrows), where $\theta(t)$ denotes the instantaneous angle between the position vector and the negative $x$ axis. If the run is long enough, the swimmer can escape as it reaches point $E$ where $\mathbf{p}$ becomes tangent with the surface.} \label{fig:collisiondyn} \vspace{-0.1cm}
\end{figure}

We now turn to the dynamics during a collision, and analyze swimmer motion after it first impacts with the pillar and still has time $\tau_r$ remaining before its next tumble.\ A schematic of a collision is shown in Fig.~\ref{fig:collisiondyn}.\ 
For the purpose of calculating the displacement $\delta\mathbf{r}$, we lose no generality by choosing a Cartesian coordinate system with the $x$ axis aligned with the current swimming direction $\mathbf{p}$ and the origin at the center of the pillar.\ 
 We denote by $C$ the position of the collision point, which forms an angle $\alpha\in[-\pi/2,\pi/2]$ with the negative $x$ axis.\ Due to the symmetry $\alpha\leftrightarrow -\alpha$, we can restrict our attention to collisions for which $\alpha\ge 0$.\ Note that the incidence angle is a random variable, whose probability density function depends on the type of collision. For a collision of type A, the normal coordinate $y_c=a\sin\alpha$ is uniformly distributed over $[-a,a]$ since the pillar location is uniformly distributed in the shaded region of Fig.~\ref{fig:collision}(a), and therefore
\begin{equation}
p_\mathrm{A}(\alpha)=\cos\alpha,\qquad \alpha\in[0,\pi/2]. \label{eq:alphadistA}
\end{equation}
However, for collisions of type B, the angle $\alpha$ itself is uniformly distributed, i.e.,
 \begin{equation}
p_\mathrm{B}(\alpha)=\frac{2}{\pi},\qquad \alpha\in[0,\pi/2]. \label{eq:alphadistB}
\end{equation}

As the swimmer moves along the pillar surface, its orientation $\mathbf{p}$ does not change by assumption. Instead, the swimmer slides with tangential velocity $v_0(\mathbf{I}-\hat{\mathbf{n}}\hat{\mathbf{n}})\cdot\mathbf{p}$, where $\hat{\mathbf{n}}$ is the unit normal on the surface.\ This translates into the angular velocity
\begin{equation}
\frac{\mathrm{d}\theta}{\mathrm{d}t}=\frac{v_0}{a}\sin\theta,
\end{equation}
where the angle $\theta(t)$ defines the angular position of the swimmer on the pillar as shown in Fig.~\ref{fig:collisiondyn}.\ 
This can be integrated as
\begin{equation}
\int_{\alpha}^\theta\frac{\mathrm{d}\theta}{\sin\theta}= \int_{0}^t \frac{v_0}{a} \,\mathrm{d}t,
\end{equation}
i.e.,
\begin{equation}
\log\left[\frac{\tan(\theta/2)}{\tan(\alpha/2)}\right]=\frac{v_0}{a}t, \label{eq:thetaevolution}
\end{equation}
where we have chosen the origin of time $t=0$ as the instant when contact first takes place: $\theta(0)=\alpha$.

\begin{figure}[t]
\includegraphics[width=\columnwidth]{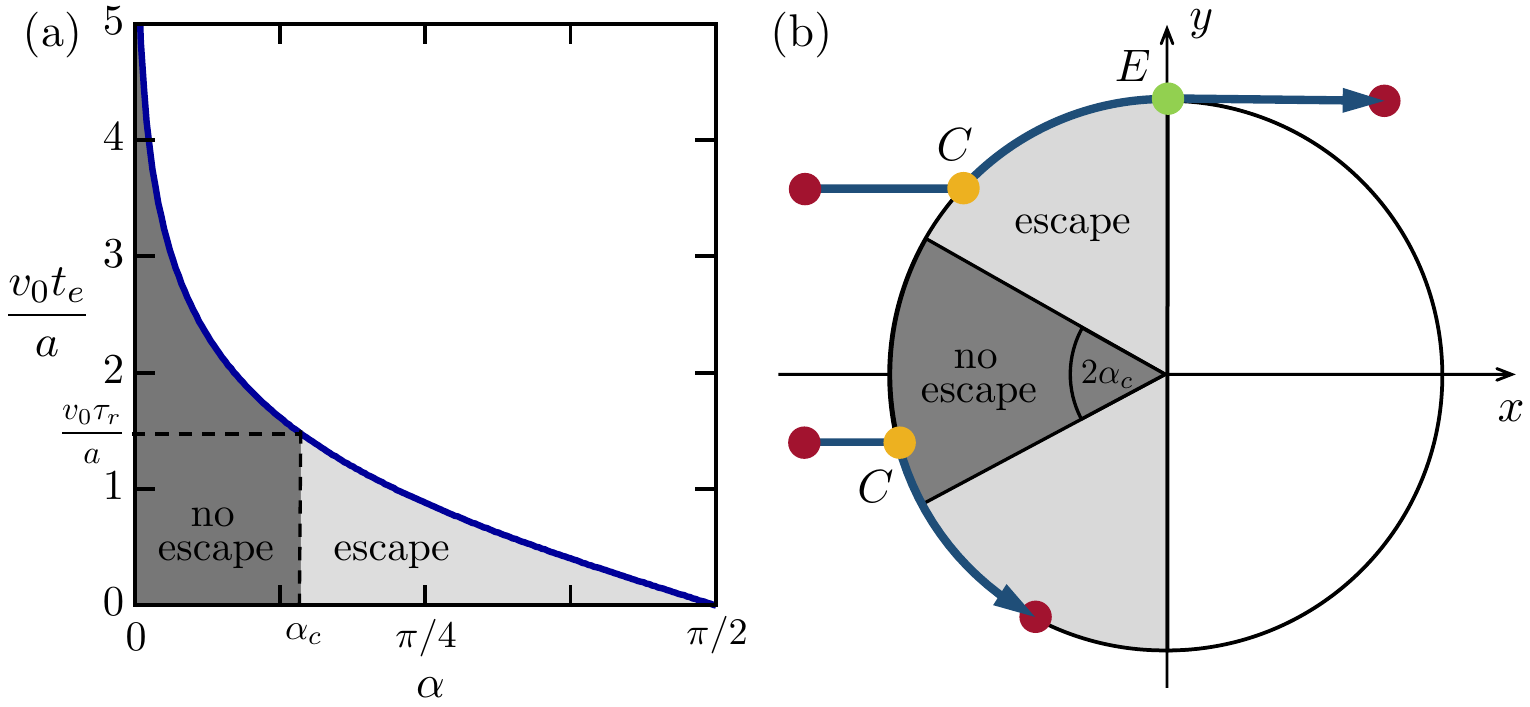}\vspace{-0.0cm}
\caption{(a) Escape time $t_e(\alpha)$ as a function of incidence angle. The swimmer will escape if $\tau_r\ge t_{e}(\alpha)$. (b) Critical angle for escape: for the swimmer to escape, its incidence angle must fall outside of a wedge of angle $2\alpha_c$. } \label{fig:escapetime}
\end{figure}

There are two possible outcomes to a collision.\ If $\theta$ reaches $\pi/2$ before the end of the run, the swimmer escapes the pillar at point $E$ in Fig.~\ref{fig:collision}(b) and finishes its run in a straight line.\ Otherwise, the current run will end at some location $\theta_{\!f}\in[\alpha,\pi/2)$ where the next tumble will take place.\ The time for the swimmer to reach $E$, or escape time $t_e$, is found by setting $\theta=\pi/2$ in Eq.~(\ref{eq:thetaevolution}):
\begin{equation}
t_{e}(\alpha)=-\frac{a}{v_0}\log \tan(\alpha/2).
\end{equation}
The escape time is plotted in Fig.~\ref{fig:escapetime}(a) and shows a strong dependence on incidence angle $\alpha$, with $t_e(\alpha)\rightarrow \infty$ as $\alpha\rightarrow 0$.  Indeed, a swimmer hitting a pillar nearly head-on ($\alpha\gtrsim 0$) initially slides very slowly as its tangential velocity goes as $\sin\alpha$, whereas a swimmer hitting a pillar nearly tangentially ($\alpha\lesssim \pi/2$) is able to escape after a short time. 

For the swimmer to escape before the end of the current run, the remaining time $\tau_r$ after contact should exceed the escape time:
\begin{equation}
\tau_r\ge t_e(\alpha). \label{eq:escapecond}
\end{equation}
For a given value of $\tau_r$, this gives a condition on the incidence angle: the swimmer will escape if $\alpha\ge \alpha_c$ where
\begin{equation}
\alpha_c(\tau_r) =2\tan^{-1} [\exp (-v_0\tau_r/a)] ,
\end{equation}
 but will finish the current run on the surface of the pillar otherwise; see Fig.~\ref{fig:escapetime}(b). If the swimmer escapes, it continues its run in the $x$ direction after leaving the surface of the pillar at point $E$, for a duration of $\tau_r-t_{e}(\alpha)$.

\begin{figure}[t]
\includegraphics[width=\columnwidth]{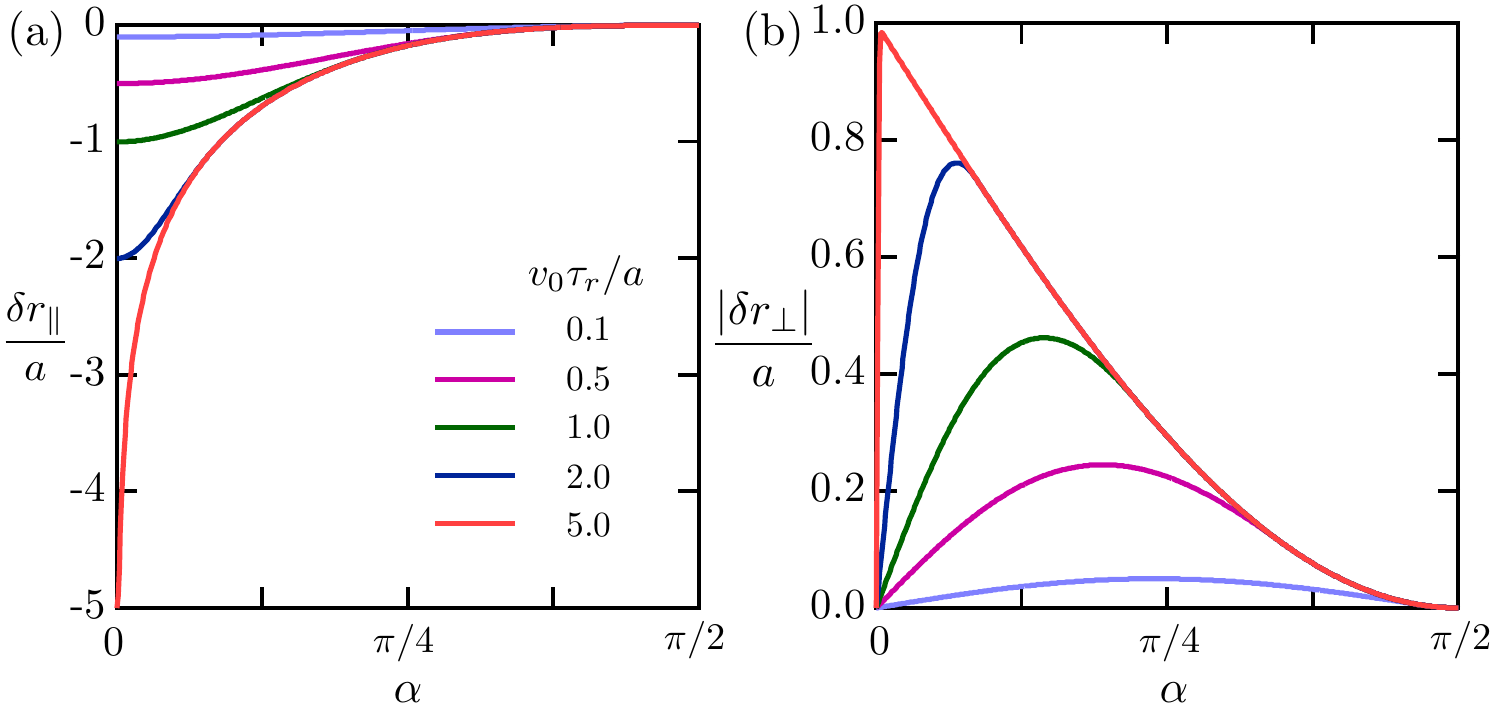}
\caption{(a) Longitudinal displacement $\delta r_\parallel/a$ and (b) transverse displacement $|\delta r_\perp |/a$ as functions of incidence angle $\alpha$, for different values of $v_0\tau_r/a$ where $\tau_r$ is the remaining time in the run after collision. } \label{fig:Deltaalpha}
\end{figure}

We can now estimate the longitudinal and transverse displacements incurred by the collision with the pillar. We first consider the case where the swimmer escapes the pillar at point $E$, i.e., $\alpha\ge\alpha_c$ or $\tau_r\ge t_e$.\ In the $x$ direction, the swimmer undergoes a displacement of $a\cos \alpha$ over the course of the collision, while it would have travelled a distance of $v_0t_e(\alpha)$ during the same amount of time, had there been no collision. Therefore,
\begin{equation}
\Delta x = a \left[\cos\alpha + \log \tan (\alpha/2)\right].
\end{equation}
In the transverse direction, the displacement is easily obtained as
\begin{equation}
\Delta y= a (1-\sin\alpha).
\end{equation}
On the other hand, if the run time elapses before the swimmer escapes, i.e., $\alpha<\alpha_c$ or $\tau_r< t_e$, the swimmer will finish the current run at angular position $\theta_{\!f}$ on the pillar surface, where
\begin{equation}
\begin{split}
\theta_{\!f}(\alpha,\tau_r)= 2 \tan^{-1}\left[\frac{\tan(\alpha/2)}{\exp(- v_0 \tau_r/a)}\right]. \label{eq:thetae}
\end{split}
\end{equation}
In the longitudinal direction, the displacement over the course of the collision is $a[\cos\alpha-\cos\theta_{\!f}]$, whereas it would have been $v_0\tau_r$ in the absence of collision. Therefore
\begin{equation}
\Delta x=a(\cos\alpha-\cos\theta_{\!f})-v_0\tau_r,
\end{equation}
while the transverse displacement is simply given by
\begin{equation}
\Delta y = a (\sin\theta_{\!f}-\sin\alpha).
\end{equation}
In summary, the longitudinal and transverse displacements incurred by a collision are expressed as
\begin{equation}
\frac{\delta r_\parallel}{a}=\begin{cases}
\cos\alpha + \log \tan (\alpha/2) & \alpha\ge \alpha_c, \\
\cos\alpha-\cos\theta_{\!f}-v_0\tau_r/a & \alpha<\alpha_c,
\end{cases} \label{eq:rpara}
\end{equation}
and
\begin{equation}
\frac{|\delta r_\perp|}{a}=\begin{cases}
 1-\sin\alpha & \alpha\ge \alpha_c, \\
 \sin\theta_{\!f}-\sin\alpha & \alpha<\alpha_c,
\end{cases} \label{eq:rperp}
\end{equation}
where $\theta_{\!f}$ is given by Eq.~(\ref{eq:thetae}). Note that $\delta r_\parallel\le 0$, whereas $r_\perp$ is of either sign by symmetry: collisions hinder longitudinal transport but induce transverse motion of either sign. The displacements $\delta r_\parallel $ and $|\delta r_\perp|$ are plotted vs incidence angle $\alpha$ in Fig.~\ref{fig:Deltaalpha}. As expected, collisions have the greatest effect on transport at vanishing incidence angles ($\alpha\rightarrow 0$), for which $\delta r_\parallel\rightarrow v_0\tau_r$ and $|\delta r_\perp|\rightarrow a$ for large $v_0\tau_r/a$.

\subsection{Probability of escape\label{sec:escape}}

\begin{figure}[t]
\includegraphics[width=0.75\columnwidth]{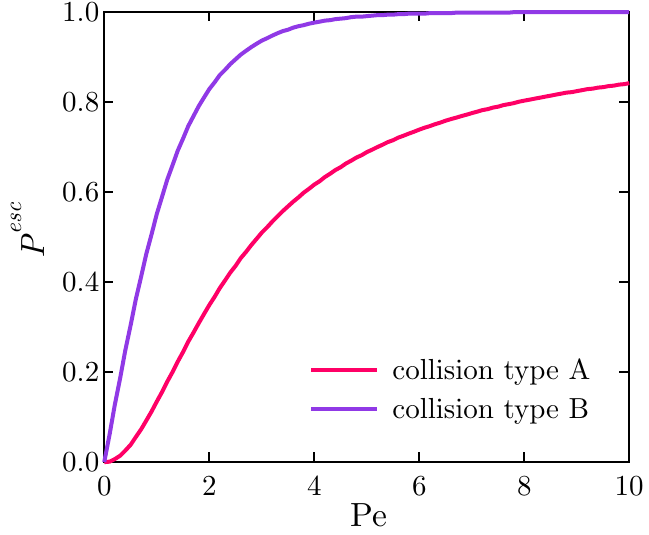}\vspace{-0.1cm}
\caption{Escape probability $P^{esc}$ for a collision of type A or B as a function of P\'eclet number, as obtained in Eqs.~(\ref{eq:escapeA})--(\ref{eq:escapeB}).} \label{fig:escapeprob}\vspace{-0.1cm}
\end{figure}

We are now in a position to calculate the escape probabilities $P^{esc}_\mathrm{A}$ and $P^{esc}_\mathrm{B}$ for each type of collision, which are needed to estimate the collision probability $P^c_\mathrm{B}$ in Eq.~(\ref{eq:Pb2}).\ For a given collision, escape will occur if the condition of Eq.~(\ref{eq:escapecond}) is met.\ Therefore, taking into account all possible incidence angles,
\begin{align}
P^{esc}&=\int_{0}^{\pi/2}\left[1-P(\tau_r\le t_e(\alpha))\right]\,p(\alpha)\,\mathrm{d}\alpha, \\
&=\int_{0}^{\pi/2}\left[1-\int_0^{t_e} p(\tau_r)\,d\tau_r\right]p(\alpha)\,\mathrm{d}\alpha.
\end{align}
Inserting the probability density functions $p(\tau_r)$ and $p(\alpha)$ for each type of collision, as provided in Eqs.~(\ref{eq:taurdistA}), (\ref{eq:taurdistB}), (\ref{eq:alphadistA}) and (\ref{eq:alphadistB}), we obtain after simplifications
\begin{align}
P^{esc}_\mathrm{A}&= 1-\frac{1}{\mathrm{Pe}}\left(\alpha_0-\frac{\pi}{2}\right), \label{eq:escapeA} \\
P^{esc}_\mathrm{B}&=1-\frac{2}{\pi}\alpha_0, \label{eq:escapeB}
\end{align}
where $\alpha_0=\alpha_c(\tau)=2\tan^{-1}[\exp (-\mathrm{Pe})]$ is the critical angle for escape for a collision with $\tau_c\rightarrow 0$.\ The two escape probabilities $P^{esc}_\mathrm{A}$ and $P^{esc}_\mathrm{B}$ only depend on the P\'eclet number and are plotted in Fig.~\ref{fig:escapeprob}. For both types of collisions, the escape probability $P^{esc}$ increases monotonically with $\mathrm{Pe}$, vanishes in the limit of short runs ($\mathrm{Pe}\rightarrow 0$) and tends to $1$ in the limit of long runs ($\mathrm{Pe}\rightarrow \infty$). Collisions of type B are more likely to lead to an escape than collisions of type A as they have maximum remaining time $\tau_r=\tau$.

\begin{figure*}[t]
\includegraphics[width=\textwidth]{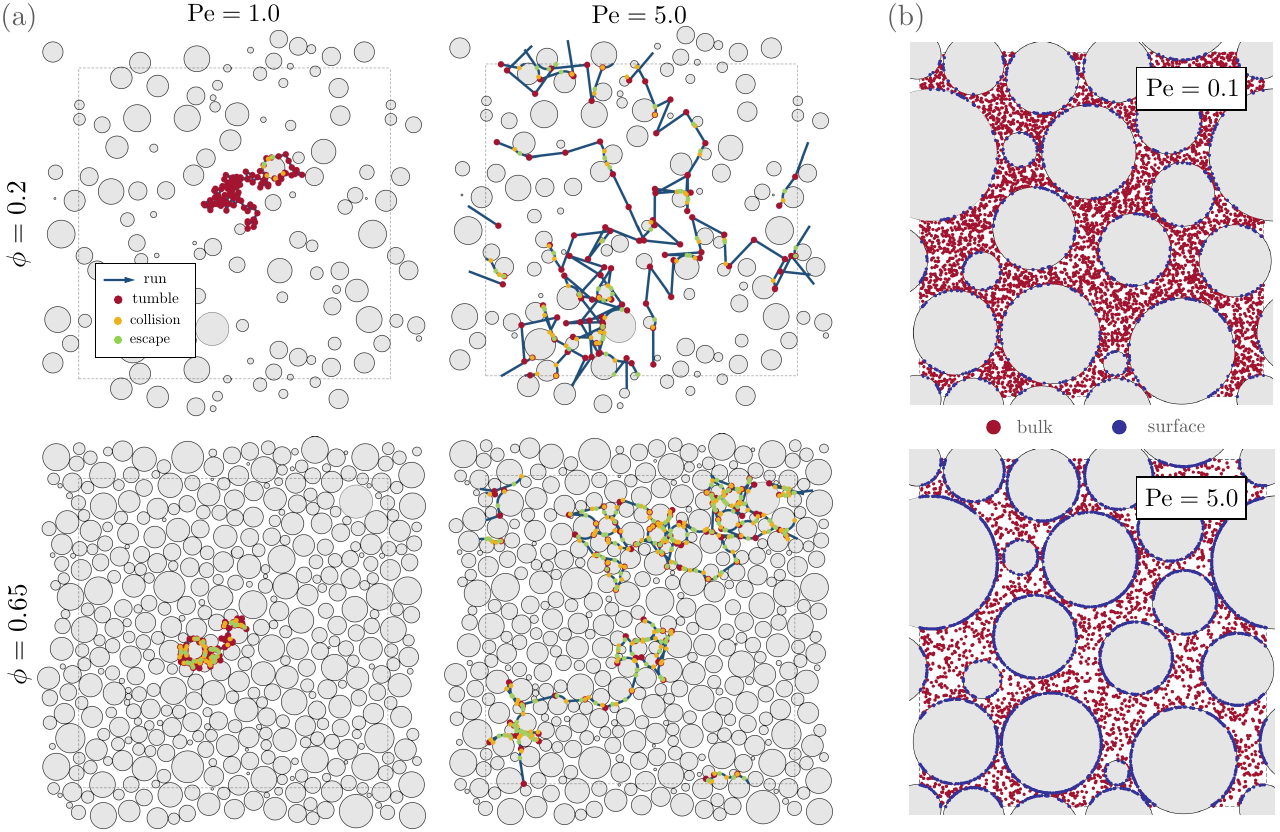}\vspace{-0.cm}
\caption{Event-based stochastic simulations in random polydisperse media. (a) Single swimmer trajectories consisting of 100 runs of constant run time, for various combinations of $\mathrm{Pe}$ (columns) and $\phi$ (rows). Red, yellow and green symbols show the location of tumbles, collisions and escape points. Also see movies in the Supplemental Material \cite{Note1}. (b) Locations of 5000 random tumbles in simulations with $\phi=0.65$ for two different values of $\mathrm{Pe}$, where the locations of tumbles occurring in the bulk or on the surface of a pillar are highlighted in red and blue, respectively. In all simulations shown, pillar radii were drawn from a Gaussian distribution with mean $\overline{a}=1$ and standard deviation $\sigma_a/\overline{a}=0.5$, and periodic boundary conditions are used at the edges of the square domain marked by a dotted line. } \label{fig:stochsim}\vspace{-0.cm}
\end{figure*}

\subsection{Displacement statistics and hindrance function}

In the case of constant run time $\tau$, the hindrance function introduced in Eq.~(\ref{eq:hindrance}) simplifies to
\begin{equation}
f(\mathrm{Pe},\phi)=- \frac{2\langle \delta r_\parallel \rangle}{\mathrm{Pe}}- \frac{\langle \delta r_\parallel^2 \rangle+\langle \delta r_\perp^2 \rangle}{\mathrm{Pe}^2}. \label{eq:hind2}
\end{equation}
We obtained analytical expressions for the displacements $\delta r_\parallel$ and $\delta r_\perp$ in Eqs.~(\ref{eq:rpara})--(\ref{eq:rperp}). The ensemble average in Eq.~(\ref{eq:hind2}) is evaluated over all possible outcomes of a run:
\begin{align}
\begin{split}
\hspace{-0.2cm}\langle\chi \rangle = P^c_\mathrm{A} \int_{0}^{\tau} \int_{0}^{\pi/2}\chi\, p_\mathrm{A}(\alpha) p_\mathrm{A}(\tau_r)\, \mathrm{d}\alpha\, \mathrm{d}\tau_r  \\
+ P^c_\mathrm{B} \int_{0}^{\tau} \int_{0}^{\pi/2}\chi\, p_\mathrm{B}(\alpha) p_\mathrm{B}(\tau_r)\, \mathrm{d}\alpha\, \mathrm{d}\tau_r ,
\end{split}\label{eq:average}
\end{align}
where the various probability density functions are given in Eqs.~(\ref{eq:taurdistA})--(\ref{eq:taurdistB}) and (\ref{eq:alphadistA})--(\ref{eq:alphadistB}). Note that $\tau=\mathrm{Pe}$ in dimensionless variables.\ The only dependence on area fraction $\phi$ in Eq.~(\ref{eq:average}) is through the prefactors of $P_\mathrm{A}^c$ and $P_\mathrm{B}^c$, which are both proportional to $1-\exp[-(2/\pi)\mathrm{Pe}\,\phi]$. 

In the limit of low volume fraction and small P\'eclet number, $\mathrm{Pe},\phi\rightarrow 0$, asymptotic expansions of the average displacements can be obtained, with leading-order contributions given by:
\begin{align}
\langle\delta_\parallel\rangle & \approx -\frac{5}{3\pi} \mathrm{Pe}^2 \phi,  \label{eq:asym1} \\
\langle\delta_\parallel^2\rangle & \approx \frac{199}{180\pi} \mathrm{Pe}^3 \phi, \\
\langle\delta_\perp^2\rangle & \approx \frac{61}{180\pi} \mathrm{Pe}^3 \phi, \label{eq:asym2}
\end{align}
from which the hindrance function is obtained as
\begin{equation}
f(\mathrm{Pe},\phi)\approx \frac{17}{9\pi}\mathrm{Pe}\,\phi.   \label{eq:asymf}
\end{equation}
At arbitrary values of $\mathrm{Pe}$ and $\phi$, the integrals in Eq.~(\ref{eq:average}) can be evaluated using numerical quadrature.\ We discuss results from this calculation in Sec.~\ref{sec:constant}, where we compare the dilute theory predictions to event-based stochastic simulations valid for a wide range of $\mathrm{Pe}$ and $\phi$.

\section{Results and discussion\label{sec:simulations}}

\subsection{Event-based stochastic simulations}

We perform event-based stochastic simulations of run-and-tumble microswimmer trajectories through randomly generated porous geometries. $N_t$ non-overlapping pillars are distributed at random inside a square periodic box to achieve the desired area fraction. The pillars can be either of uniform size or polydisperse (see Sec.~\ref{sec:polydisp}). The simulations track the positions of non-interacting run-and-tumble swimmers whose kinematics follow the assumptions of Sec.~\ref{sec:probdef}. At the start of each run, the next run time and a new random orientation are selected, potential collisions are detected, and the swimmer position is advanced until the end of the run, where the location of potential collision and escape points is obtained analytically based on the calculations of Sec~\ref{sec:dyncol}. Multiple collisions can occur during one run. For each swimmer trajectory, the simulation records the times and locations of all tumbles, collisions and escape points. The simulation box is typically chosen to be significantly larger than the mean run length, so that the statistics are unaffected by the periodic boundary conditions. 

\begin{figure}[t]
\includegraphics[width=0.9\columnwidth]{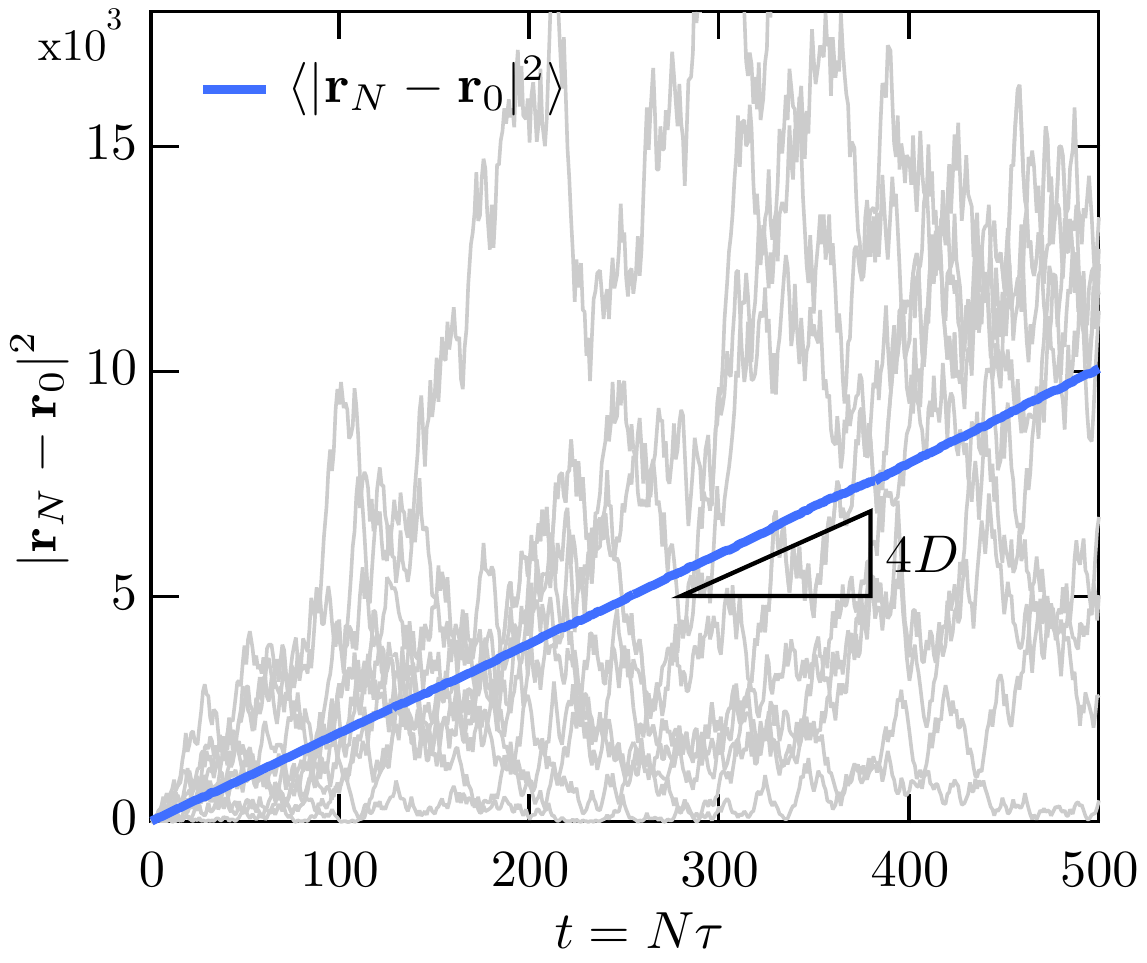}\vspace{-0.cm}
\caption{Mean squared displacement as a function of time in a typical simulation with uniform pillar size and constant run time. Gray curves show the square displacement $|\mathbf{r}_N-\mathbf{r}_0|^2$ for 10 individual stochastic simulations with distinct random seeds. The blue curve the mean squared displacement $\langle |\mathbf{r}_N-\mathbf{r}_0|^2\rangle$ obtained as an average over 1000 trajectories. A linear fit is used to obtain the diffusivity $D$ as the quarter slope. } \label{fig:msd} \vspace{-0.cm}
\end{figure}

Typical trajectories showing the locations of these points in simulations with constant run time but varying pillar size are plotted in Fig.~\ref{fig:stochsim}(a) for different combinations of P\'eclet number and area fraction (also see movies in the Supplemental Material \cite{Note1}). Expectedly, the most efficient dispersion occurs in dilute media at large $\mathrm{Pe}$ (long runs that are largely unimpeded by the medium), and increasing area fraction strongly hinders dispersion for all P\'eclet numbers.  As $\mathrm{Pe}$ increases, the swimmers spend a greater fraction of their time sliding on the surface of pillars. This is illustrated in Fig.~\ref{fig:stochsim}(b), showing the locations of 5000 tumbles for two values of $\mathrm{Pe}$: as P\'eclet number increases and swimmer trajectories become more persistent, a larger fraction of tumbles occurs on the surface of pillars. We quantify some of these trends further in the following sections. The calculation of the diffusivity from simulation data is illustrated in Fig.~\ref{fig:msd}, showing the growth of the mean squared displacement for 10 individual trajectories, as well as an average over an ensemble of 1000 trajectories.

\subsection{Constant run time and pillar size \label{sec:constant}}

We center the following discussion on results in systems with constant run time and uniform pillar size, which are the assumptions of the theoretical model of Sec.~\ref{sec:dilutetheory}. The effects of variable run time and pillar size will be briefly considered in numerical simulations in Sec.~\ref{sec:polydisp}.

\subsubsection{Collision probabilities}

\begin{figure}[t]
\includegraphics[width=0.95\columnwidth]{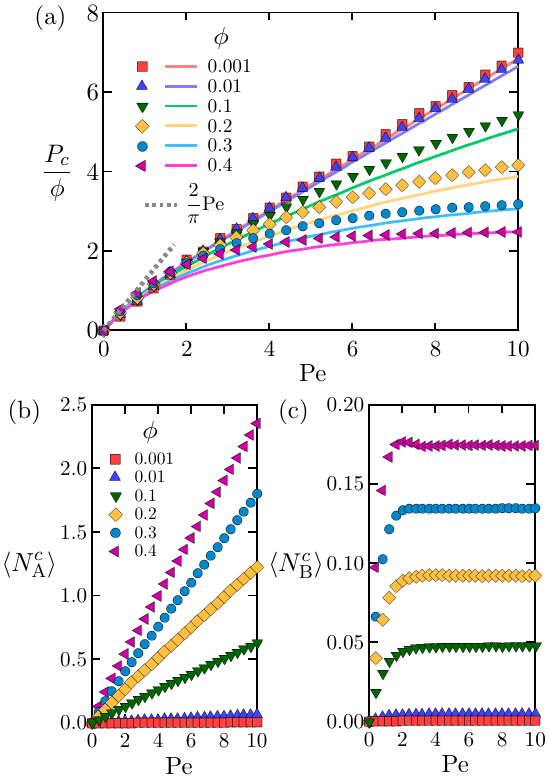}\vspace{-0.cm}
\caption{(a) Probability $P_c$ of having at least one collision (of either type A or B) within a given run, scaled by $\phi$ and plotted as a function of P\'eclet number for various area fractions. Symbols show results from stochastic simulations with uniform pillar size and constant run time, and lines show the theoretical prediction of Eq.~(\ref{eq:Pc}). (b)--(c) Mean numbers of collisions of type A (b) or type B (c) in any given run as functions of P\'eclet number for various area fractions, from stochastic simulations.} \label{fig:contactprob}\vspace{-0.cm}
\end{figure}

\begin{figure*}[t]
\includegraphics[width=\textwidth]{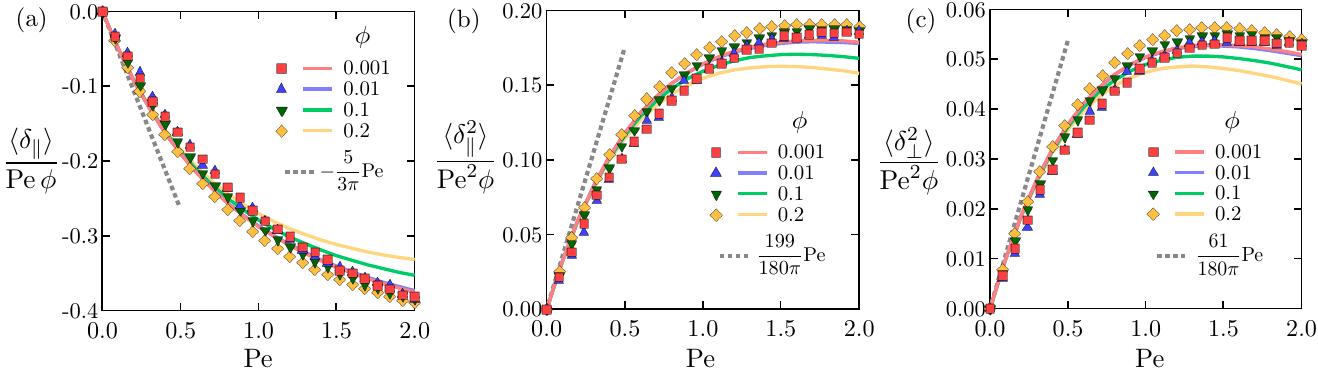}\vspace{0.cm}
\caption{Displacement statistics at low area fraction and P\'eclet number: (a) $\langle \delta_\parallel\rangle/\mathrm{Pe}\,\phi$, (b) $\langle \delta_\parallel^2\rangle/\mathrm{Pe}^2 \phi$, and (c) $\langle \delta_\perp^2\rangle/\mathrm{Pe}^2 \phi$, plotted as functions of P\'eclet number for various area fractions $\phi$. In each case, symbols show results from stochastic simulations with uniform pillar size and constant run time, whereas full lines show theoretical predictions based on the dilute theory of Sec.~\ref{sec:dilutetheory}. Dotted grey lines show the theoretical asymptotes of Eqs.~(\ref{eq:asym1})--(\ref{eq:asym2}) in the limit of $\mathrm{Pe},\phi\rightarrow 0$.  }\vspace{0.1cm} \label{fig:displacements}
\end{figure*}

We first analyze collision probabilities in Fig.~\ref{fig:contactprob}, where we compare results from stochastic simulations with theoretical predictions.\ Figure~\ref{fig:contactprob}(a) shows the probability $P_c=P^c_\mathrm{A}+P^c_\mathrm{B}$ of having at least one collision (of either type A or B) within a given run.\ The dilute theory of Sec.~\ref{sec:dilutetheory} provides the expression
\begin{equation}
P_c=\frac{2-P^{esc}_\mathrm{A}+P^{esc}_\mathrm{B}}{1+P^{esc}_\mathrm{B}}\left[1-\exp \left(-\frac{2}{\pi}\mathrm{Pe}\,\phi\right)\right],   \label{eq:Pc}
\end{equation}
where the escape probabilities $P^{esc}_\mathrm{A}$ and $P^{esc}_\mathrm{B}$ are functions of $\mathrm{Pe}$ only and were obtained in Eqs.~(\ref{eq:escapeA})--(\ref{eq:escapeB}). Remarkably, the dilute theory provides an excellent quantitative estimate of $P_c$ over a wide range of area fractions and P\'eclet numbers, well beyond its expected range of validity. In very sparse media ($\phi\ll 1$), the collision probability $P_c$ increases linearly with both $\phi$ and $\mathrm{Pe}$, while it is found to saturate with respect to $\mathrm{Pe}$ in denser media. In the limit of $\mathrm{Pe}\rightarrow \infty$, every run will incur at least one collision, so that $P_c\rightarrow 1$. 

Note that while the dilute theory assumes that at most one collision can take place during one run, such is not the case in simulations. To quantify this further, we plot in Fig.~\ref{fig:contactprob}(b,c) the mean numbers $\langle N^c_\mathrm{A}\rangle$ and $\langle N^c_\mathrm{B}\rangle$ of collisions of type A and B in any given run, from stochastic simulations. Multiple collisions of type A can occur in a run, especially in dense media at high P\'eclet numbers. Indeed, we find that $\langle N^c_\mathrm{A}\rangle$ increases nearly linearly with both $\phi$ and $\mathrm{Pe}$, and exceeds 1 at sufficiently large values of either $\phi$ or $\mathrm{Pe}$. We expect the dilute theory of Sec.~\ref{sec:dilutetheory} to be inaccurate in those regimes, since it assumes that at most one collision occurs per run. On the other hand, there cannot be more than one collision of type B in a given run: $N^c_\mathrm{B}\in \left\{ 0,1\right\}$ and therefore $\langle N^c_\mathrm{B}\rangle <1$ as seen in Fig.~\ref{fig:contactprob}(c). For all area fractions, $\langle N^c_\mathrm{B}\rangle$ first increases with $\mathrm{Pe}$ to reach a plateau for $\mathrm{Pe}\gtrsim 2$, with the value of the plateau displaying a linear dependence on $\phi$. 



\subsubsection{Displacement statistics and hindrance function \label{sec:disps}}

Next, we turn to displacement statistics, focusing on the limit of low area fraction and P\'eclet number.\ Figure~\ref{fig:displacements} shows the relevant statistics entering the calculation of the hindrance function in Eq.~(\ref{eq:hind2}) as functions of P\'eclet number for various area fractions: panel (a) shows the mean longitudinal  displacement $\smash{\langle \delta_\parallel\rangle}$ scaled by $\mathrm{Pe}\,\phi$, whereas panels (b) and (c) show the variances of the longitudinal and transverse displacements, $\smash{\langle \delta_\parallel^2\rangle}$ and $\smash{\langle \delta_\perp^2\rangle}$, respectively, both scaled by $\mathrm{Pe}^2\phi$.\ At low P\'eclet number, all the displacements collapse and are very well captured by the asymptotic results of Eqs.~(\ref{eq:asym1})--(\ref{eq:asym2}), which predict a linear dependence on $\phi$, as well as a linear dependence on $\mathrm{Pe}$ upon rescaling.\ As the P\'eclet number is increased, the growth of the displacement statistics with $\mathrm{Pe}$ slows down and ultimately saturates, yet the collapse with respect to area fraction persists. The dilute theory of Sec.~\ref{sec:dilutetheory} is found to provide excellent quantitative predictions for $\phi\lesssim 0.01$ over the range of P\'eclet numbers considered here. Departures are observed at larger volume fractions when $\mathrm{Pe}\gtrsim 1$, beyond which the dilute theory underpredicts displacements: this can be attributed to the fact that the dilute theory assumes at most one collision per run, whereas multiple collisions of type A typically occur in that regime in simulations, as previously found in Fig.~\ref{fig:contactprob}(b). Finally, we note that the magnitude of $\smash{\langle \delta_\perp^2\rangle/\mathrm{Pe}^2\phi}$ is notably smaller than $\smash{\langle \delta_\parallel \rangle/\mathrm{Pe}\,\phi}$ and $\smash{\langle \delta_\parallel^2\rangle/\mathrm{Pe}^2\phi}$, indicating that the leading contribution to the hindrance function comes from the reduction in longitudinal displacements. 

\begin{figure}[t]\vspace{-0.0cm}
\includegraphics[width=0.85\columnwidth]{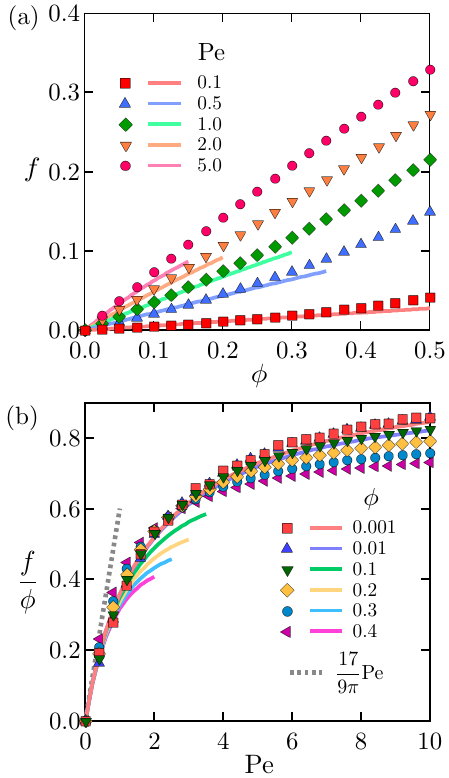}\vspace{-0.cm}
\caption{(a) Hindrance function $f$ as a function of area function $\phi$ for various values of P\'eclet number $\mathrm{Pe}$. (b) Hindrance function $f$, scaled by $\phi$, as a function of $\mathrm{Pe}$ for various values of $\phi$. In both panels, symbols show results from stochastic simulations with uniform pillar size and constant run time, and full lines show theoretical predictions from the dilute theory of Sec.~\ref{sec:dilutetheory}. Dotted line in (b) shows the low-Pe asymptote of Eq.~(\ref{eq:asymf}).  } \label{fig:hindrance}\vspace{-0.cm}
\end{figure}

The hindrance function $f(\mathrm{Pe},\phi)$ is analyzed in Fig.~\ref{fig:hindrance}, where we compare results from stochastic simulations (symbols) with the predictions from the dilute theory (lines).\ The dependence on area fraction is shown in Fig.~\ref{fig:hindrance}(a), showing $f$ as a function of $\phi$ for various P\'eclet numbers. The hindrance is found to grow nearly linearly with $\phi$ for all values of $\mathrm{Pe}$ considered here, as expected from the collapse of the displacement statistics upon scaling by $\phi$ in Fig.~\ref{fig:displacements}. Good agreement with the theoretical prediction is observed, especially at low $\phi$ and $\mathrm{Pe}$, consistent with the assumptions of the theory; departures are observed as $\phi$ increases, where the theory systematically underpredicts the hindrance function. The dependence on P\'eclet number is illustrated in Fig.~\ref{fig:hindrance}(b), where we show $f$ scaled by $\phi$ as a function of $\mathrm{Pe}$. At low P\'eclet number, the simulation data matches the theoretical model very well and collapses onto the asymptotic prediction of Eq.~(\ref{eq:asymf}), which predicts a linear dependence on $\mathrm{Pe}$.\ Upon increasing the P\'eclet number, the growth of $f/\phi$ slows down and ultimately saturates, reaching a plateau whose value depends weakly on $\phi$, with larger values attained at lower area fractions. Consistent with the observations in Fig.~\ref{fig:displacements}, the dilute theory for the hindrance function is found to provide an excellent fit to the data in dilute media ($\phi\lesssim 0.01$) even when the P\'eclet number is large, but it significantly underpredicts $f$ at larger values of $\phi$, due to the preponderance of runs with multiple collisions. 

\subsection{Variable run time and pillar size \label{sec:polydisp}}

\begin{figure}[t]\vspace{-0.0cm}
\includegraphics[width=0.85\columnwidth]{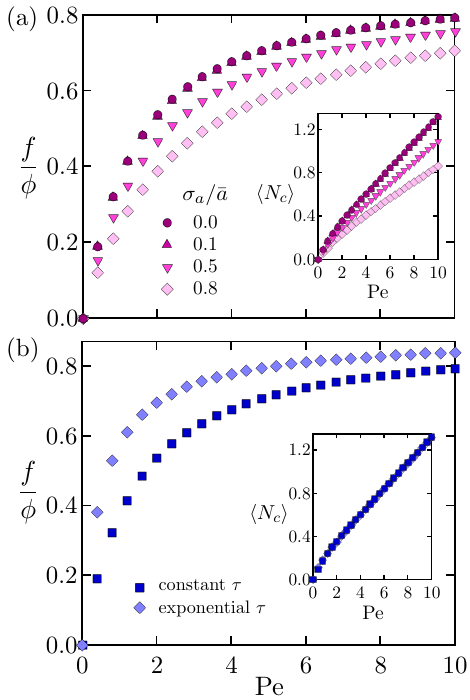}\vspace{-0.cm}
\caption{Scaled hindrance function $f/\phi$ as a function of P\'eclet number, for: (a) simulations with constant run time in several polydisperse media, where 
$\sigma_a$ is the standard deviation of the pillar radius distribution; and (b) simulations with uniform pillar size and either constant or exponentially distributed run times.\ In both panels, $\phi=0.2$.\ The insets show the average number of collisions of any type per run, $\langle N_c\rangle=\langle N^c_\mathrm{A}+N^c_\mathrm{B}\rangle$, for the same conditions.  } \label{fig:variable}\vspace{-0.cm}
\end{figure}

The previous results have exclusively considered the case of constant run time and monodisperse media---two assumptions that are convenient for theoretical analysis but unlikely to be met in many experimental systems of interest.\ Here, we relax these assumptions and analyze the effects of varying run time and pillar size using stochastic simulations.  

We first consider the effect of obstacle polydispersity on the hindrance function in Fig.~\ref{fig:variable}(a). Porous media of increasing polydispersity were generated by drawing pillar radii from Gaussian distributions of increasing widths (while rejecting negative values). The generated distributions were then rescaled affinely to have mean 1, and their measured standard deviations $\sigma_a$ are reported in the figure. Weak polydispersity ($\sigma_a/a=0.1$) has only a negligible effect on dispersion.\ The hindrance function, however, is reduced by up to $\sim20\%$ in highly polydisperse media ($\sigma_a/a=0.5$ and $0.8$), with the strongest effect occurring for intermediate P\'eclet numbers ($\mathrm{Pe}\sim 2-6$). That dispersion is easier in a polydisperse medium is, perhaps, an intuitive result, for the same reason that it is easier to pack polydisperse particles than monodisperse ones. The decrease in $f$ can simply be explained by a decrease in the mean number of collisions per run, $\langle N_c\rangle=\langle N^c_\mathrm{A}+N^c_\mathrm{B}\rangle$, as polydispersity becomes significant; see inset of Fig.~\ref{fig:variable}(a). 

The effect of variable $\tau$ is analyzed in Fig.~\ref{fig:variable}(b), comparing the hindrance function for constant and exponentially distributed run times, in a system with uniform pillars and $\phi=0.2$. In this case, variations in run time cause an increase in the value of $f$, especially at low to intermediate P\'eclet numbers ($\mathrm{Pe}\sim 1-4$).\ The reason for this difference is less intuitive: indeed, the mean number of collisions per run is nearly unaffected by variations in run time, as shown in the inset.\ Instead, we attribute it to a change in the relative magnitude of the averages appearing in Eq.~(\ref{eq:hindrance}), and the effect on the hindrance is most pronounced at low P\'eclet numbers, where the displacement statistics are most sensitive to variations in~$\mathrm{Pe}$. 

\section{Concluding remarks\label{sec:conclusion}}

We have presented a minimal theoretical model for the dispersion of run-and-tumble microswimmers in disordered porous media composed of randomly distributed circular pillars in two dimensions.\ The effect of the microstructure on the long-time spatial dispersion was shown to be entirely captured by a scalar dimensionless hindrance function $f(\phi,\mathrm{Pe})$ of the medium area fraction $\phi$ and swimming P\'eclet number $\mathrm{Pe}$, which compares the persistence length of swimmer trajectories to the size of the solid inclusions.\ Under simple assumptions for the interaction of the microswimmers with the microstructure, we were able to obtain an analytical expression for the hindrance function in the dilute limit of $\mathrm{Pe}\,\phi\ll 1$, and stochastic simulations were performed to extend this result to the case of denser media. The hindrance function was shown to depend nearly linearly on area fraction over a wide range of parameter values---an intuitive result since the number of collisions incurred during a run increases linearly with $\phi$. The dependence on P\'eclet number was also found to be linear at low values of $\mathrm{Pe}$, but to saturate at larger values of $\mathrm{Pe}$. While the analytical prediction captured the data very well for $\mathrm{Pe}\lesssim O(1)$, it was found to underestimate the hindrance function at moderate to high P\'eclet numbers in relatively dense media, where multiple collisions can occur during a given run. Because of its relative simplicity and ease of analysis, the framework proposed here provides a basis for the interpretation and analysis of experimental data and for the benchmarking of more complex models. 

We emphasize that the model we developed here relies on strong simplifying assumptions that may not be satisfied in many experimental systems.\ We only considered two-dimensional systems composed of circular non-overlapping pillars:\ while such geometries have indeed been analyzed in microfluidic experiments \cite{CCDDA2019,BBCPR2019,DWDG2019,DWG2023}, natural disordered media typically involve three-dimensional microstructures that are significantly more complex. Extending our model to three dimensions is tedious but relatively straightforward; allowing for overlapping or non-circular occlusions, however, is significantly more involved and unlikely to be tractable analytically. The role of obstacle shape is expected to be of particular interest: non-convex obstacles may indeed result in trapping of microswimmers with a strong effect on dispersion \cite{VV2017}, whereas asymmetric shapes can induce a net drift by a rectification mechanism \cite{DZTALRWSZ2017,ACS2019}.\ Note also that our model assumed point-sized microswimmers, which are able to pass through arbitrarily thin gaps. In reality, finite-sized swimmers may get trapped when attempting to travel through thin gaps, forcing them to reverse direction as has been observed in experiments on bacteria in dense media \cite{BD2019,BD2019b}; accounting for this motility strategy requires distinct modeling choices \cite{PBDPS2021,KMBLDS2021} easily incorporated in a framework such as ours. 

Another major assumption of our model is that of frictionless sliding during collisions, with no change to the swimmer orientation.\ In particular, this assumes that interactions are purely steric and that hydrodynamic effects are negligible. Experiments on various systems have shown that hydrodynamic interactions can reorient and trap microswimmers near circular obstacles \cite{TPBSG2014,SNDG2015}, as can chemical interactions in the case of self-phoretic particles \cite{DGCHSVGE2015,SKUPTS2016}.\ Other types of active particles, e.g., Quincke rollers, may also undergo more complex scattering dynamics \cite{MCCB2017}. Accounting for such effects in our model is possible in principle. Understanding the role of external fields, such as applied flows \cite{CCDDA2019,DWDG2019} or chemical gradients \cite{APYSJ2021,BAAOD2022}, is also an open problem of great interest, which would require solving for the local velocity or chemical field in the porous matrix, for instance using the boundary element method. Finally we note that our model has focused on the transport of dilute non-interacting swimmer suspensions: the case of semi-dilute to dense suspensions, which can undergo spontaneous flow transitions in confinement \cite{TAS2017}, has been considered in a few experimental \cite{WWDG2016,NASS2018,RNHSBKA2020} and computational \cite{TS2019} studies in periodic porous media, but remains an open area of investigation. Some of these open questions will be addressed in future work. \vspace*{0.2cm}

\begin{acknowledgments} \vspace*{-0.1cm}
The author thanks Can Yang and Antoine Beringer for help with preliminary simulations, and Tanumoy Dhar for useful conversations. This work was funded by National Science Foundation Grant CBET--1934199.
\end{acknowledgments}

\bibliography{RTPrefs}

\end{document}